\documentstyle[aps,epsf,preprint]{revtex}
\tightenlines
\begin{document}

\title{Analysis of nature of \boldmath{$\phi\to\gamma\pi\eta$} and
\boldmath{$\phi\to\gamma\pi^0\pi^0$} decays
\thanks{Plenary session talk at HADRON 2001, Protvino, Russia, August 27}}
\author{N.N. Achasov\\
Laboratory of Theoretical Physics,
 Sobolev Institute for Mathematics\\
  Academician Koptiug prospekt, 4,
 630090 Novosibirsk, Russia}
 \maketitle

\begin{abstract}
 We study interference patterns in the $\phi\to(\gamma
a_0+\pi^0\rho)\to\gamma\pi\eta$ and
 $\phi\to(\gamma f_0+\pi^0\rho)\to\gamma \pi^0\pi^0$ reactions.
 Taking into account the interference,  we fit the
 experimental data and show that the background reaction
 does not distort the $\pi^0\eta$ spectrum in the decay
$\phi\to\gamma\pi\eta$ everywhere over the energy region
 and does not
 distort the $\pi^0\pi^0$ spectrum in the decay
$\phi\to\gamma\pi^0\pi^0$ when the invariant mass
$m_{\pi^0\pi^0}>670$ MeV. We discuss the details of the scalar
meson production in the radiative decays and note that there are
conclusive arguments in favor of the one-loop mechanism $\phi\to
K^+K^-\to\gamma a_0\,(\, \mbox{or}\ \gamma f_0\, )$.  We discuss
also distinctions between the four-quark, molecular, and two-quark
models and argue that the establishment of the scalar meson
production mechanism in the $\phi$ radiative decays gives new
strong evidence in favor of the four-quark nature of the scalar
$a_0(980)$ and $f_0(980)$ mesons.
\end{abstract}
\vspace*{1.5cm}

\section{Introduction}

As  was shown in a number of papers, see Refs.
\cite{achasov-89,close-93,nutral,shevchenko,lucio,phase} and
references therein, the study of the radiative decays
$\phi\to\gamma a_0\to\gamma\pi\eta$ and $\phi\to\gamma f_0\to
\gamma\pi\pi$ can shed light on the problem of the scalar
$a_0(980)$ and $f_0(980)$ mesons. These decays have been studied
not only theoretically but also experimentally \cite{misha}.
Present time data have already been obtained from Novosibirsk with
the detectors SND
 \cite{snd-1,snd-2,snd-fit,snd-ivan} and CMD-2 \cite{cmd},
 which give the following
branching ratios :
$BR(\phi\to\gamma\pi\eta)=(0.88\pm0.14\pm0.09)\cdot10^{-4}$
\cite{snd-fit}, $BR(\phi\to\gamma\pi^0\pi^0)=
(1.221\pm0.098\pm0.061)\cdot10^{-4}$ \cite{snd-ivan} and
$BR(\phi\to\gamma\pi\eta)=(0.9\pm0.24\pm0.1)\cdot10^{-4}$,
$BR(\phi\to\gamma\pi^0\pi^0)=(0.92\pm0.08\pm0.06)\cdot10^{-4}$
\cite{cmd}. DA$\Phi$NE also confirms the Novosibirsk results
\cite{barbara}.

 These data give evidence in favor of the four-quark $(q^2\bar
q^2)$
\cite{achasov-89,jaffe,ach-84,ach-91,ach-98,black,achasov-01,josef,tuan,tesh}
nature of the scalar $a_0(980)$ and $f_0(980)$ mesons. Note that
the isovector $a_0(980)$ meson is produced in the radiative $\phi$
meson decay
 as intensively as the well-studied $\eta'$ meson involving
 essentially strange quarks $s\bar s$
($\approx66\%$), responsible for the decay.

As shown in Refs. \cite{achasov-89,nutral,bramon}, the background
situation for studying the radiative decays $\phi\to\gamma
a_0\to\gamma\pi^0\eta$ and $\phi\to\gamma f_0\to \gamma\pi^0\pi^0$
is very good. For example, in the case of the decay $\phi\to\gamma
a_0\to\gamma\pi^0\eta$, the process
$\phi\to\pi^0\rho\to\gamma\pi^0\eta$ is the dominant background.
The estimation for the soft, by strong interaction standard,
photon energy, $\omega<100$ MeV, gives
$BR(\phi\to\pi^0\rho^0\to\gamma\pi^0\eta,\omega<100\ \mbox{MeV})
\approx 1.5\cdot10^{-6}$. The influence of the background process
is negligible, provided
 $BR(\phi\to\gamma a_0\to\gamma\pi^0\eta,\omega<100\ \mbox{MeV})\geq10^{-5}$.
 In Sec. II
we show that for the obtained experimental data the influence of
the background processes is negligible everywhere over the photon
energy region \cite{achasov-01}.

The situation with  $\phi\to\gamma f_0\to \gamma
 \pi^0\pi^0$ decay is not much different. As was shown in
\cite{achasov-89,nutral,bramon} the dominant background is the
$\phi\to\pi^0\rho^0\to\gamma\pi^0\pi^0$ process with
$BR(\phi\to\pi^0\rho^0\to\gamma\pi^0\pi^0,\omega<100\ \mbox{MeV})
\approx 6.4\cdot10^{-7}$. The influence of this background process
is negligible, provided $BR(\phi\to\gamma
f_0\to\gamma\pi^0\pi^0,\omega<100\ \mbox{MeV})\geq5\cdot10^{-6}$.

 The exact calculation of the interference patterns between the decays
$\phi\to\gamma f_0\to \gamma\pi^0\pi^0$ and
$\phi\to\rho^0\pi\to\gamma\pi^0\pi^0$ \cite{achasov-01}, which we
present
 in Sec. III, shows that the influence of the
background in the decay
 $\phi\to\gamma\pi^0\pi^0$ for the obtained experimental data
 is negligible in the wide region of
 the $\pi^0\pi^0$ invariant mass, $m_{\pi\pi}>670$ MeV,
 or in the photon energy region $\omega<300$ MeV.

 In Sec. IV we discuss the mechanism of the scalar meson production in the
 radiative decays and show that experimental data obtained in Novosibirsk
give the conclusive arguments in favor of the one-loop mechanism
$\phi\to K^+K^-\to\gamma a_0$ and $\phi\to K^+K^-\to\gamma f_0$ of
these decays \cite{achasov-01}. We explain also why this
circumstance gives new strong evidence  in favor of the four-quark
nature of the scalar $a_0(980)$ and $f_0(980)$ mesons.

\section{Interference between the reactions \lowercase{\boldmath{$\phi\to\gamma
a_0\to\gamma \pi^0\eta$}} and
\lowercase{\boldmath{$\phi\to\pi^0\rho^0\to\gamma\pi^0\eta$}}}

\begin{figure}
\centerline{\epsfxsize=14cm \epsfysize=9.5cm \epsfbox{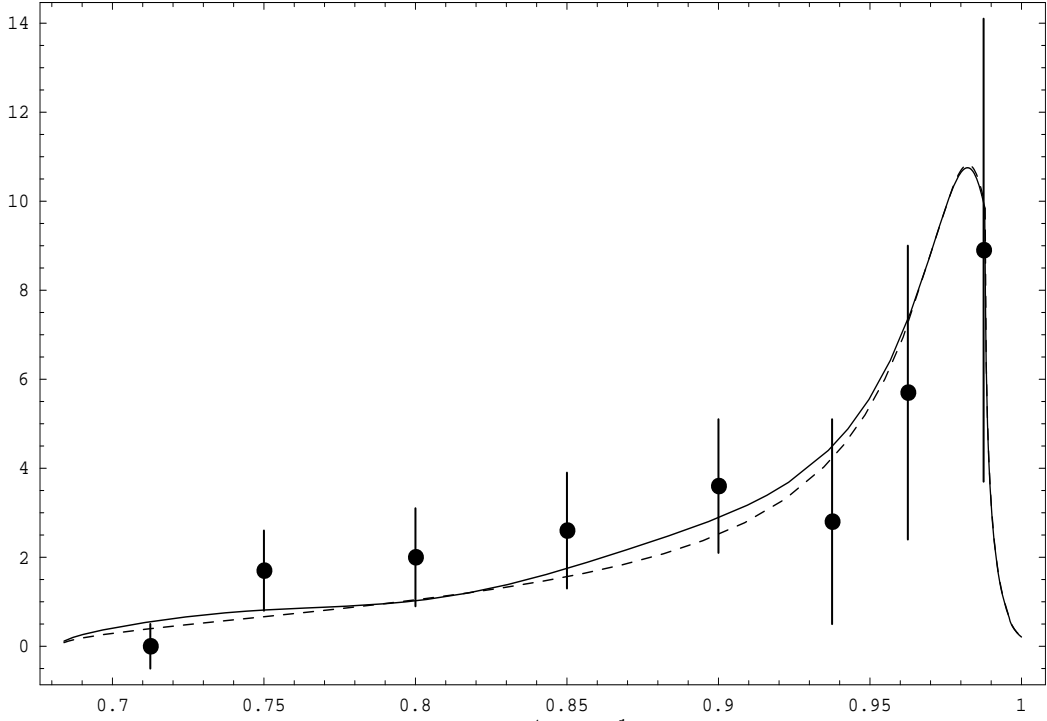}}
 \caption{ Fitting of $dBR(\phi\to\gamma\pi\eta)/dm\times 10^4\mbox{GeV}^{-1}$  with
the background is shown with the solid line, the signal
contribution is shown with the dashed line.}  \label{figa0}
\end{figure}

 The fit \cite{achasov-01} of the experimental data from the SND detector
\cite{snd-fit} is shown in Fig. \ref{figa0}.
The total branching
ratio, taking into account the interference, is $BR(\phi\to(\gamma
a_0+\pi^0\rho)\to\gamma\pi\eta)=(0.79\pm0.2)\cdot10^{-4}$, the
branching ratio of the signal is  $BR(\phi\to\gamma
a_0\to\gamma\pi\eta)=(0.75\pm0.2)\cdot10^{-4}$ and the branching
ratio of the background is
$BR(\phi\to\rho^0\pi^0\to\gamma\pi^0\eta)=3.43\cdot10^{-6}$. So,
the integral part of the interference is negligible. The influence
of the interference on the mass spectrum of the $\pi\eta$ system
is also negligible, see Fig. \ref{figa0}.

\section{Interference between the
\lowercase{\boldmath{$e^+e^-\to\gamma f_0\to\gamma \pi^0\pi^0$}}
and
\lowercase{\boldmath{$e^+e^-\to\phi\to\pi^0\rho\to\gamma\pi^0\pi^0$}}
reactions}

 The total branching ratio, with interference being taken into
 account, is
  $BR(\phi\to(\gamma
f_0+\pi^0\rho)\to\gamma\pi^0\pi^0)=(1.26\pm0.29)\cdot10^{-4}$, the
branching ratio of the signal is $BR(\phi\to\gamma
f_0\to\gamma\pi^0\pi^0)=(1.01\pm0.23)\cdot10^{-4}$,
 the branching ratio of the background is
$BR(\phi\to\rho^0\pi^0\to\gamma\pi^0\pi^0)=0.18\cdot10^{-4}$.

\begin{figure}
\centerline{\epsfxsize=14cm \epsfysize=9.5cm
\epsfbox{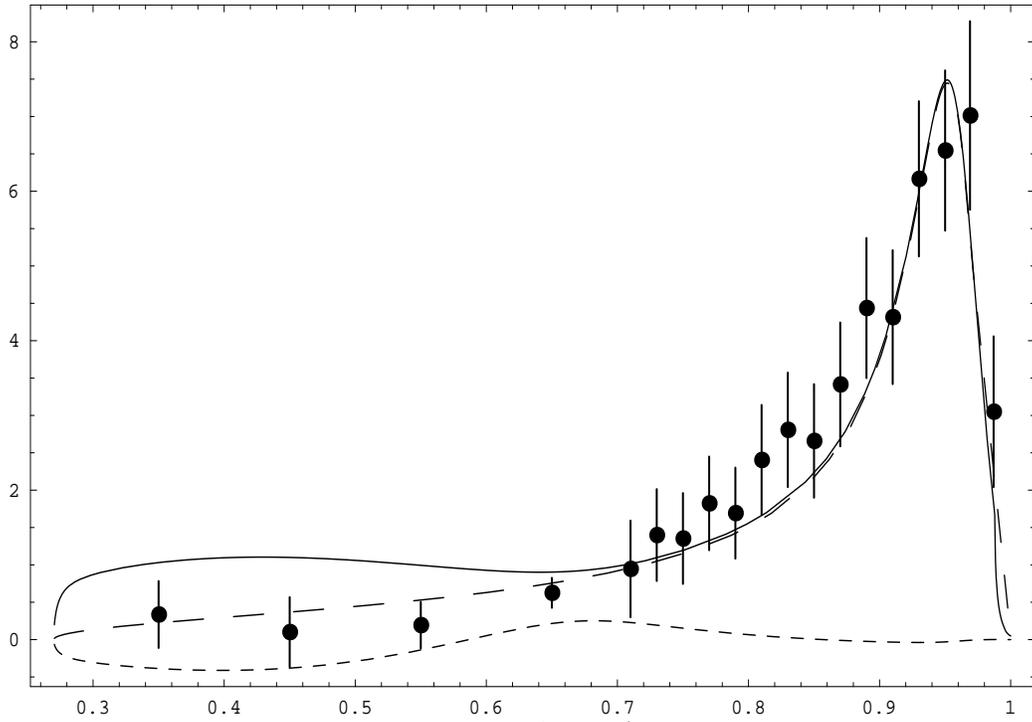}}
 \caption{
Fitting of $dBR(\phi\to\gamma\pi^0\pi^0)/dm\times
10^{4}\mbox{GeV}^{-1}$
  with the background is shown with the
solid line, the signal contribution is shown with the dashed line.
The dotted line is the interference term.
 The data are from the SND detector.
} \label{figf0snd}
\end{figure}

One can see from  Fig.  \ref{figf0snd} that the influence of the
background process on the spectrum of the
$\phi\to\gamma\pi^0\pi^0$ decay  is negligible in the wide region
of the $\pi^0\pi^0$ invariant mass, $m_{\pi\pi}>670$ MeV, or when
photon energy less than $300$ MeV.

The difference from the experimental data,
 observed in the
region $m_{\pi\pi}<670$ MeV, is due to the fact that in the
experimental processing of the $e^+e^-\to\gamma\pi^0\pi^0$ events
the background events $e^+e^-\to\omega\pi^0\to\gamma\pi^0\pi^0$
are excluded with the help of the invariant mass cutting and
simulation, in so doing the part of the
$e^+e^-\to\phi\to\rho\pi^0\to\gamma\pi^0\pi^0$ events is excluded
as well.

\section{Conclusion}

The experimental data give evidence \cite{achasov-01} not only in
favor of the four-quark model but in favor of the dynamical model
 suggested in Ref. \cite{achasov-89}, in which
the discussed decays proceed through the kaon loop, $\phi\to K^+
K^- \to\gamma f_0(a_0)$, see Refs. \cite{achasov-89,nutral}

%see Fig. \ref{model}.
%\begin{figure}
%\centerline{\epsfxsize=14cm \epsfysize= 4cm \epsfbox{model.eps}}
%\caption{Diagrams of the $K^+K^-$ loop model.} \label{model}
%\end{figure}

\begin{figure}
\centerline{\epsfxsize=14cm \epsfysize=10cm \epsfbox{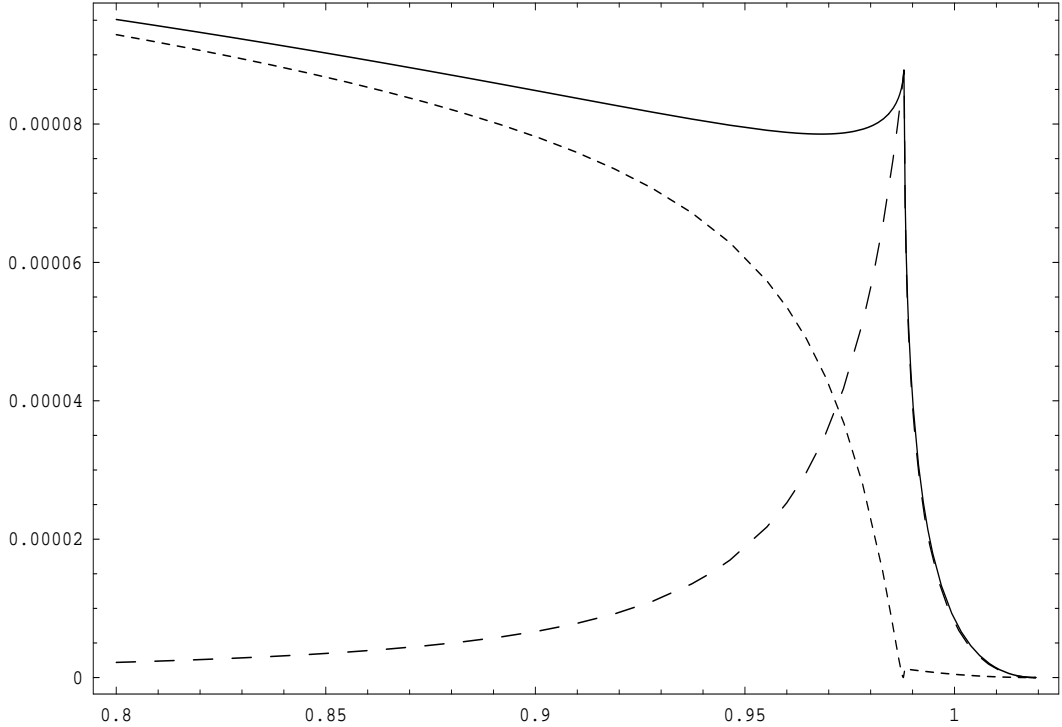}}
 \caption{ The function $|g(m)|^2$ is drawn with the solid line. The contribution of the
  imaginary part is drawn with the dashed line. The contribution of the real part
   is drawn with the dotted line.} \label{g}
\end{figure}

Indeed, according to the gauge invariance condition, the
transition amplitude $\phi\to\gamma f_0(a_0)$ is proportional to
the electromagnetic field strength tensor $F_{\mu\nu}$ (in our
case to the electric field). Since there are no pole terms in our
case, the function $g(m)$ in
\begin{eqnarray}
&&\frac{d\Gamma(\phi\to\gamma a_0\to\gamma\pi\eta\,m)}{dm}=
\frac{2}{\pi}\frac{m^2\Gamma(\phi\to\gamma
a_0(m))\Gamma(a_0(m)\to\pi\eta)}{|D_{a_0}(m)|^2}\nonumber\\[1pc]
&&= \frac{2|g(m)|^2p_{\eta\pi}(m_{\phi}^2-m^2)}
{3(4\pi)^3m_{\phi}^3}\left
|\frac{g_{a_0K^+K^-}g_{a_0\pi\eta}}{D_{a_0}(m)}\right |^2
\label{spectruma0}
\end{eqnarray}
and \footnote{Eq. (\ref{f0}) takes into account the mixig of
$f_0(980)$ meson with other scalar resonances, see Refs.
\cite{nutral,achasov-01}.}
\begin{eqnarray}
&&\frac{d\Gamma(\phi\to\gamma f_0\to\gamma\pi^0\pi^0\,m)
}{dm}\nonumber\\[1pc]
&&=\frac{|g(m)|^2\sqrt{m^2-4m_{\pi}^2}(m_{\phi}^2-m^2)}
{3(4\pi)^3m_{\phi}^3}\left
|\sum_{R,R'}g_{RK^+K^-}G_{RR'}^{-1}g_{R'\pi^0\pi^0}\right |^2
\label{f0}
\end{eqnarray}
is proportional to the energy of photon
$\omega=(m_{\phi}^2-m^2)/2m_{\phi}$ in the soft photon region. To
describe the experimental spectra in Figs. \ref{figa0} and
\ref{figf0snd}, the function $|g(m)|^2$ should be smooth (almost
constant) in the range $m\leq0.99$ GeV, see Eqs.
(\ref{spectruma0}) and (\ref{f0}). Stopping the function
$\omega^2$ at $\omega_0=30$ MeV, using the form-factor of the form
$1/(1+R^2\omega^2)$, requires $R\approx 100$ GeV$^{-1}$. It seems
to be incredible to explain the formation of such a huge radius in
hadron physics. Based on the large, by hadron physics standard,
$R\approx10$ GeV$^{-1}$, one can obtain an effective maximum of
the mass spectra under discussion only near 900 MeV. In the
meantime, the $K^+K^-$ loop
%, see Fig. \ref{model},
  gives the
natural description to this threshold effect, see Fig. \ref{g}.

\begin{figure}
\centerline{\epsfxsize=14cm \epsfysize=8.5cm \epsfbox{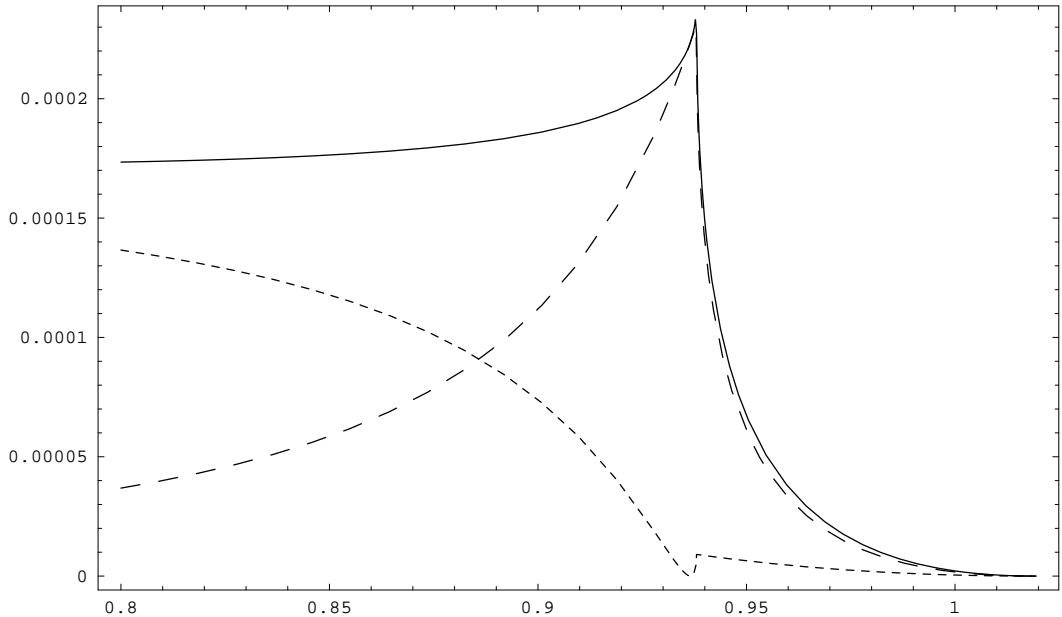}}
 \caption{ The function $|g(m)|^2$ for  $m_{K^+}=469$ MeV is drawn
 with the solid line. The contribution of the imaginary  is drawn with the dashed line.
 The contribution of the real part is drawn with the dotted line.}
  \label{gg}
\end{figure}

\begin{figure}
\centerline{ \epsfxsize=14cm \epsfysize=8.5cm \epsfbox{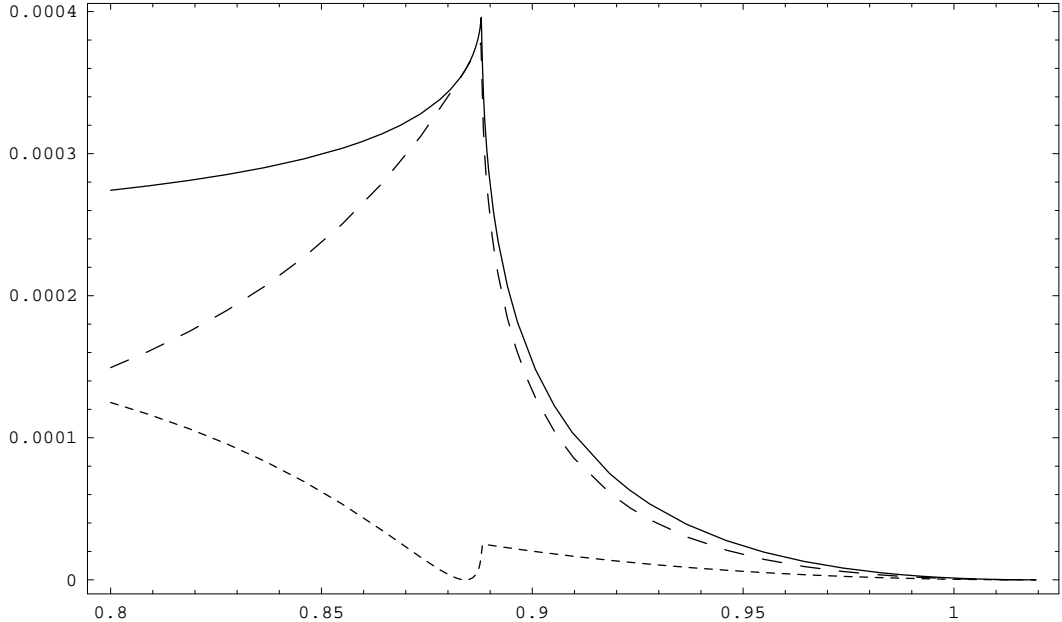}}
 \caption{ The function $|g(m)|^2$ for  $m_{K^+}=444$ MeV is drawn
  with the solid line. The contribution of the imaginary part is drawn with the dashed line.
 The contribution of the real part is drawn with the dotted line.
 } \label{ggg}
\end{figure}

 To demonstrate the threshold character of this effect we present
 Fig. \ref{gg} and Fig.\ref{ggg} in which the function $|g(m)|^2$
is shown in the case of $K^+$ meson mass is 25 MeV and 50 MeV less
than in reality.

One can see from Figs. \ref{gg} and \ref{ggg} that the function
$|g(m)|^2$ is suppressed by the $\omega^2$ low in the region
950-1020 MeV and 900-1020 Mev respectively.

In the mass spectrum this suppression is increased by one more
power of $\omega$, see Eqs. (\ref{spectruma0}) and (\ref{f0}), so
that we cannot see the resonance in the region 980-995 MeV. The
maximum in the spectrum is effectively shifted to the region
935-950 MeV and 880-900 MeV respectively.

In truth this means that $a_0(980)$ and $f_0(980)$ resonances are
seen in the radiative decays of $\phi$ meson owing to the $K^+K^-$
intermediate state, otherwise the maxima in the spectra would be
shifted to 900 MeV.

Thus the mechanism of production of the scalar mesons in the
$\phi$ radiative decays is established at a physical level of
proof. It is the rarest case in hadron physics. \footnote{It is
worth noting that the $K^+K^-$ loop model is practically accepted
by theorists, compare, for example, Ref. \cite{marco} with Ref.
\cite{br}; true there is exception \cite{anis}.}

Both real and imaginary parts of the $\phi\to\gamma a_0(f_0)$
amplitudes are caused by the $K^+K^-$ intermediate state. The
imaginary parts are caused by the real $K^+K^-$ intermediate state
while the real parts are caused by the virtual compact $K^+K^-$
intermediate state, that is, we are dealing here with the
four-quark transition \footnote{It will be recalled that the
imaginary part of every hadronic amplitude describes a multi-quark
transition.}. Needless to say radiative four-quark transitions can
happen between two two-quark states as well as between two-quark
and four-quark states but their intensities depend strongly on a
type of the transitions. A radiative four-quark transition between
two two-quark states requires creation of an additional $q\bar q$
pair, that is, such a transition is forbidden according to the
Okuba-Zweig-Izuka (OZI) rule, while a radiative four-quark
transition between two-quark and four-quark states happens without
creation an additional $q\bar q$ pair, that is, such a transition
is allowed according to the OZI rule.

Let us discuss this problem from two point of views: i) from
 point of view of intermediate states and ii) from point of view of the $1/N_c$
expansion .

i) It was noted already in paper \cite{achasov-89} that the
imaginary part of the $K^+K^-$  loop is calculated practically in
a model independent way making use of the coupling constants
$g_{\phi K^+K^-}$ and $g_{a_0(f_0)K^+K^-}$ due to the Low's
theorem \cite{low} for the photons with energy $\omega<100$ MeV
which is soft by the standard of strong interaction.

In the same paper it was noted that the real part of the loop
(with accuracy up to 20\% in the width of the $\phi\to\gamma
f_0(a_0)$ decay) is practically not different for the point-like
particle and the compact hadron with form-factor which has the
cutting radius in the momentum space about the mass of  $\rho$
meson ($m_{\rho}=0.77$ GeV).

In contrast to the four-quark state which is the compact hadron
\cite{jaffe}, the bound $K\bar K$ state is the extended state with
the spatial radius $R\sim1/\sqrt{m_K\epsilon}$, where $\epsilon$
is the binding energy. Corresponding form-factor in the momentum
space has the radius of the order of
$\sqrt{m_K\epsilon}\approx100$ MeV for $\epsilon=20$ MeV,
\cite{markushin}. The more detailed calculation \cite{close-93}
gives for the radius in the momentum space the value $p_0=140$
MeV. As a result, the contribution of the virtual intermediate
$K^+K^-$ states in the $K^+K^-$ loop is suppressed by the momentum
distribution in the molecule, and the real part of the loop
amplitude is negligible \cite{shevchenko}. It leads to the
branching ratio much less  than the experimental one.
%, as it was noted above.
In addition, the spectrum is much narrower in the $K\bar K$
molecule case  that contradicts to the experiment, see the
behavior of the imaginary part contribution in Fig. \ref{g} and in
corresponding figures in \cite{shevchenko}.

 Of course,  the two-quark state is as compact as four-quark one.
The question arises, why is the branching ratio in the two-quark
model suppressed in comparison with the branching ratio in the
four-quark model? There are two reasons. First, the coupling
constant of two-quark states with the $K\bar K$ channel is
noticeably less \cite{nutral,ach-84} and, second, there is the OZI
rule that is more important really.

If the isovector $a_0(980)$ meson is the two-quark state,  it has
no strange quarks. Hence \cite{achasov-89,nutral,ach-98}, the
decay $\phi\to\gamma a_0$ should be suppressed according to the
OZI rule. On the intermediate state level, the OZI rule is
formulated as compensation of the different intermediate states
\cite{lipkin,geiger,ach-kozh}. In our case these states are $K\bar
K$, $K\bar K^*+\bar KK^*$, $K^*\bar K^*$ and so on. Since, due to
the kinematical reason, the real intermediate state is the only
 $K^+K^-$ state, the compensation in the imaginary part is
 impossible and it destroys the OZI rule. The compensation should
 be in the real part of the amplitude only. As a result, the $\phi\to\gamma a_0$ decay
  in the two-quark  model is mainly due to the imaginary part of the
 amplitude and is much less intensive than in the four-quark
 model \cite{achasov-89,nutral}. In addition, in the two-quark model, $a_0(980)$ meson should
 appear in the  $\phi\to\gamma a_0$ decay as a noticeably more narrow
 resonance than in other processes, see the
behavior of the imaginary part contribution in Fig. \ref{g}.

 As regards to the isoscalar $f_0(980)$ state, there are two
possibilities  in the two-quark model. If $f_0(980)$ meson does
not contain the strange quarks  the all above mentioned arguments
about suppression of the  $\phi\to\gamma a_0$ decay and the
spectrum shape  are also valid for the  $\phi\to\gamma f_0$ decay.
Generally speaking, there could be the strong OZI violation for
the isoscalar $q\bar q$ states ( mixing of the $u\bar u$, $d\bar
d$ and $s\bar s$ states) with regard to the strong mixing of the
quark and gluon degree of freedom which is due to the
nonperturbative effects of QCD \cite{vainshtein}. But, an almost
exact degeneration of the masses of the isoscalar $f_0(980)$ and
isovector $a_0(980)$ mesons excludes  such a possibility. Note
also, the experiment points directly to the weak coupling of
$f_0(980)$ meson with gluons, $B(J/\psi\to\gamma
f_0\to\gamma\pi\pi)<1.4\cdot10^{-5}$ \cite{eigen}.

If $f_0(980)$ meson is close to the $s\bar s$ state
\cite{ach-98,tornqvist}, there is  no suppression due to the the
OZI rule. Nevertheless, if $a_0(980)$ and $f_0(980)$ mesons are
the members of the same multiplet, the $\phi\to\gamma f_0$
branching ratio,
$BR(\phi\to\gamma\pi^0\pi^0)=(1/3)BR(\phi\to\gamma\pi\pi)
 \approx1.8\cdot10^{-5}$,
  is  significantly less than that in the four-quark
model,  due to the relation between the coupling constants with
the $K\bar K$, $\pi\eta$ and $K\bar K$, $\pi\pi$ channels
inherited in the two-quark model, see Refs.
\cite{achasov-89,nutral}. In addition, in this case there is no
natural explanation of the $a_0$ and $f_0$ mass degeneration.

Only in the case when the nature of $f_0(980)$ meson in no way
related to the nature of $a_0(980)$ meson (which, for example, is
the four-quark state) the experimentally observed branching ratio
of the $\phi\to\gamma f_0$ decay could be explained   by $s\bar s$
nature of $f_0(980)$ meson. But, from the theoretical point of
view, such a possibility seems awful \cite{ach-98}.

ii) What is more,  the OZI allowed transition is bound to have a
small weight in the $1/N_c$ expansion in  case of $s\bar s$ nature
of $f_0(980)$ meson. Indeed, the main term of the $1/N_c$
expansion of the $\phi\to\gamma f_0$ amplitude, i.e, the OZI
allowed transition, has the order of $N_c^0$ but does not contains
the $K^+K^-$ intermediate state. This state emerges only in the
next to leading term of the $1/N_c$ expansion, i.e., in the OZI
forbidden transition, which has the order of $1/N_c$.

If $f_0(980)$ meson is the two-quark state without the strange
quarks, the $1/N_c$ expansion of the $\phi\to\gamma f_0(980)$
amplitude starts with the OZI forbidden transition of the order of
$1/N_c$. But a weight of this term is bound to be small, because
it does not contain the $K^+K^-$ intermediate state, which emerges
only in the next to leading term of the order of $1/N_c^2$.

In the two-quark model of $a_0(980)$ meson the $1/N_c$ expansion
of the $\phi\to\gamma a_0(980)$ amplitude starts also with the OZI
forbidden transition of the order of $1/N_c$, whose weight is
bound to be small, because this term does not contain  the
$K^+K^-$ intermediate state, which emerges only in the next to
leading term of the order of $1/N_c^2$ too \footnote{It will be
recalled that the OZI allowed $\phi\to\gamma\eta^\prime$ amplitude
has the order of $N_c^0$.}.

In the meantime, if $a_0(980)$ and $f_0(980)$ mesons are compact
$K\bar K$ states, i.e., four-quark states, the $1/N_c$ expansions
of the $\phi\to\gamma a_0(980)(f_0(980))$ amplitudes start  with
the OZI allowed transitions of the order of $N_c^{-1/2}$, which
contain the $K^+K^-$ intermediate state.

As we see, the knowledge of the mechanism of the scalar meson
production in the $\phi$ radiative decays gives the new very
strong (if not crucial) evidence in favor of the four-quark nature
of the scalar $a_0(980)$ and $f_0(980)$ mesons.

\section{Acknowledgement}

I thank very much Organizers for the invitation and the complete
financial support.\\
I gratefully acknowledge the V.V. Gubin help too.\\
This work was supported in part by INTAS-RFBR, grant
 IR-97-232.


\begin{references}
\bibitem{markushin}
Unfortunately, in the interesting paper V.E. Markushin, Eur.
Phys.J., A8, 389 (2000), the potential in the momentum space was
taken as the momentum distribution in the molecule instead of the
wave function in the momentum space. But the momentum distribution
radius of the potential is  5-8 times as large as one of the wave
function, that was the reason for the misleading conclusion on the
possibility to explain the Novosibirsk reasults in the molecule
case.
\end{references}
\end{document}